# Preparation of $Cu_xCe_{0.3-x}Ni_{0.7}Fe_2O_4$ ferrite nanoparticles as a nitrogen dioxide gas sensor

Shaymaa A. Kadhim [1] and Tagreed M. Al- Saadi [2]
[1,2]College of Education for Pure Science/Ibn Al- Haitham, University of Baghdad, Baghdad, Iraq
[1]Shaimaa.Ahmed1104a@ihcoedu.uobaghdad.edu.iq

Abstract: In this work, the ferrite nanocomposite $Cu_xCe_{0.3-x}Ni_{0.7}Fe_2O_4$ is prepared (where: x = 0, 0.05, 0.1, 0.15, 0.2, 0.25) was prepared using the auto combustion technique (sol-gel), and citric acid was utilized as the fuel for Auto combustion. The results of X-ray diffraction (XRD), emitting field scanning electron microscope (FE-SEM), and energy dispersive X-ray analyzer (EDX) tests revealed that the prepared compound has a face-centered cubic structure (FCC) polycrystalline, and the lattice constant increases with an increase in the percentage of doping for the copper ion, and decreases for the cerium ion and that the compound is porous, and its molecules are spherical, and there are no additional elements present other than those used in the synthesis of the compound, indicating that it is of high purity, and the combination has a high sensitivity to Nitrogen dioxide ($NO_2$) gas, as determined by the gas detecting equipment.

## 1. Introduction:

The need for sensors for hazardous gases including CO, $CO_2$, $NO_2$, $H_2S$, and others has seen growing attention as environmental pollution concerns and awareness of the need to monitor hazardous gases grow. A variety of solid-state device sensors for gases have been developed as a result of the demand for their detection and monitoring [1]. The popularity of semiconductor-based chemical sensors is due to their compact size, straightforward operation, high sensitivity, selectivity, and reasonably straightforward auxiliary electronics [2]. However, the lack of selectivity is a disadvantage when detecting a target gas in a mixture of gases. One of the most prevalent methods of increasing sensor selectivity and sensitivity is doping with different compounds [1, 3]. The use of various unique semiconductor oxides as sensing components for gas sensing in bulk ceramics, thick films, and thin-film forms has been investigated [4]. Spinel ferrites have been employed as an alternative material in the gas sensor business. Several sensors will be considered before focusing on ferrites as gas sensors, their crystal structure, and manufacturing processes. Gas sensors detect changes in the electrical, acoustic, visual, mass, or calorimetric characteristics of a substance. Because it is simple, quick, and inexpensive, detection based on variations in electrical characteristics is receiving the most attention. The need for sensors to be integrated into smart devices for remote sensing is increasing, and portability and operating system compatibility are hastening the development of electrical detection-based sensors [5, 6]. It has been discovered that $MFe_2O_4$ type spinel semiconductor oxides are a sensitive formula. Substances react with both oxidizing and reducing gases [7]. Researchers are interested in nano-ferrites because of their readily adjustable properties and wide range of potential applications in sensors, microwave devices, magnetic recording, adsorbents, and data storage. Hundreds of metal oxide materials are utilized as active layers in the spinel ferrite structure of gas sensors as thick or extremely thin films for the cation site. Spinel is made up of 32 oxygen atoms arranged in a cubic crystal form with 64 tetrahedral sites. 32-site octahedral gas sensors can help with chemical management, home security, and environmental monitoring. More new materials are being explored for high-performance solid-state gas sensors [8]. The ferrite's structural, electrical, and magnetic properties are impacted by the Fe-Fe reactions. [9]. Finally, the purpose of this work is to investigate the effect of replacing Ce ion with Cu on the structural and sensitivity to $NO_2$ gas features of (sol-gel) generated ferrite nanoparticles $Cu_xCe_{0.3-x}Ni_{0.7}Fe_2O_4$.

## 2. Experimental:

Auto combustion (sol-gel) was employed to prepare the raw materials for utilization. $Cu_xCe_{0.3-x}Ni_{0.7}Fe_2O_4$. The chemicals collected are listed in Table (1).

Table (1). Masses of raw materials used in the preparation of ferrite nanocomposite samples $Cu_xCe_{0.3-x}Ni_{0.7}Fe_2O_4$ and their molar ratios.

| citric acid | | cerium nitrate | | copper nitrate | | nickel nitrate | | iron nitrate | |
|---|---|---|---|---|---|---|---|---|---|
| n | m(g) | $n_{0.3-x}$ | m(g) | $n_x$ | m(g) | n | m(g) | n | m(g) |
| 3 | 23.0556 | 0.3 | 5.21075 | 0 | 0 | 0.7 | 8.14268 | 2 | 32.32 |
| 3 | 23.0556 | 0.25 | 4.3423 | 0.05 | 0.4832 | 0.7 | 8.14268 | 2 | 32.32 |
| 3 | 23.0556 | 0.2 | 3.47384 | 0.1 | 0.9664 | 0.7 | 8.14268 | 2 | 32.32 |
| 3 | 23.0556 | 0.15 | 2.60538 | 0.15 | 1.4496 | 0.7 | 8.14268 | 2 | 32.32 |
| 3 | 23.0556 | 0.1 | 1.73692 | 0.2 | 1.9328 | 0.7 | 8.14268 | 2 | 32.32 |
| 3 | 23.0556 | 0.05 | 0.86846 | 0.25 | 2.416 | 0.7 | 8.14268 | 2 | 32.32 |

In a 1000 mL heat-resistant glass beaker, the metal nitrate was added to 40 ml of water. In a separate beaker, the citric acid was added, then (40 ml) of deionized water, and finally the acid solution to the nitrate solution. The two solutions are thoroughly blended without heating using a magnetic stirring device, and then a little amount of ammonia in the form of drops was added to the mixture until the pH is equalized to (7). The magnetic stirrer heater is then activated until the mixture reaches about (90°C). The mixing process was continued with heating until the mixture became a gel, at which point the stirrer motor is turned off, while the heating continued until the gel ignites automatically and entirely. The resultant ferrite is then allowed to cool before being ground with a mortar. The nano ferrite powder is then placed in the oven for two hours for each sample. The structural characteristics of the resultant ferrite powder are then examined using XRD, FE-SEM, and EDX methods. Following the measurements, 1.5 g of powder from each sample was collected and physically pressed for one minute at 200 bar pressure, resulting in a disc with a diameter of 1 cm and a thickness of 3.5 mm for each sample. After two hours in the oven at 900°C, six samples of the synthesized ferrite were generated. The gas sensitivity method was then used to make electrodes for each sample and assess their sensitivity to $NO_2$ gas.

## 3. Results and Discussion:

### 3.1. X-Ray Diffraction:

Figure (1) represents the X-ray diffraction patterns of $Cu_xCe_{0.3-x}Ni_{0.7}Fe_2O_4$ nano ferrite samples that match the $NiFe_2O_4$ diffraction pattern of JCPDS Standard Card No.(10-0325).

The X-ray diffraction patterns of the prepared ferrites samples show more than seven clear peaks within the range (20°-80°) belonging to the surfaces: (111), (220), (311), (400), (422), (511), and (440), and the apparent peaks indicate the nature of the crystal structure of the prepared compound's ferrite powder, which was of the face-centered cubic spinel-type (FCC) [10].

The lattice parameters were determined using the "Fullprof suit software," and the crystallite size was calculated using the "Scherrer equation" [11, 12].

$D_{Sh} = K \lambda / \beta \cos\theta$ ……………………………………………………… (1)

Where: $\lambda$ the wavelength of X-ray (1.54 Å), $\beta$ the full width at half maximum, and $\theta$ the incident angle.

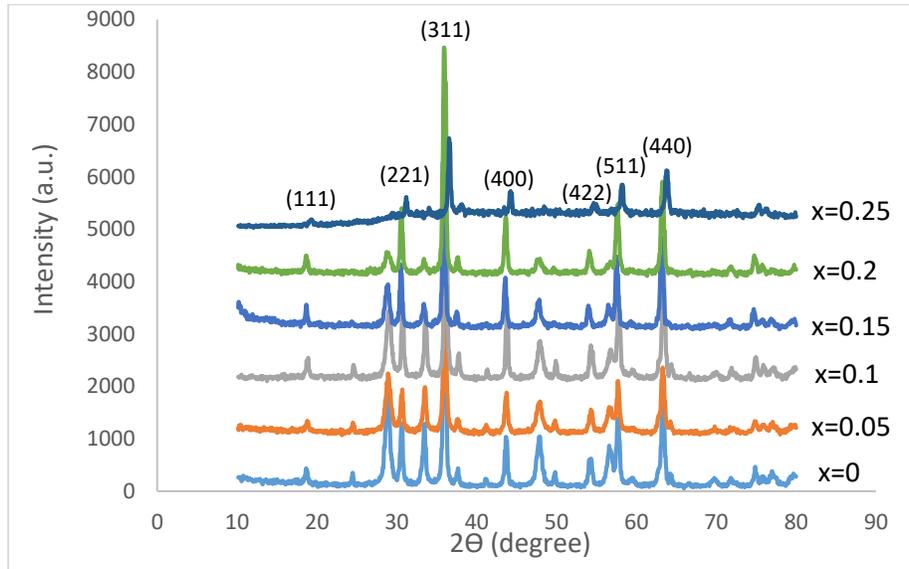

Figure (1). The X-ray diffraction patterns of $Cu_xCe_{0.3-x}Ni_{0.7}Fe_2O_4$ ferrite nanocomposite samples.

Table (2). Lattice constants (lattice constant, crystallite size and density).

| Cu Content (mole) | Lattice constant (Å) | Crystallite size (nm) | Density (g/cm³) |
|---|---|---|---|
| 0 | 8.33375 | 26.35101482 | 5.314 |
| 0.05 | 8.34517 | 30.41050846 | 5.292 |
| 0.1 | 8.34678 | 29.47659557 | 5.289 |
| 0.15 | 8.35238 | 31.86366219 | 5.278 |
| 0.2 | 8.34467 | 28.51289329 | 5.293 |
| 0.25 | 8.35222 | 23.46597741 | 5.279 |

By comparing the data in Table (2) to the molar concentrations of copper ion (x = 0, 0.05, 0.1, 0.15, 0.2, 0.25, 0.3) with the exception of one reading in Table (2), it can observe that the lattice constant increases as the proportion of copper ion increases. The addition of dopants and their increasing ratio raises the value of the lattice constant due to the migration of iron cations $Fe^{+3}$ from the tetrahedral spaces to the octahedral spaces to be replaced by impurity cations and the widening of the tetrahedral spaces as a consequence of the additional impurities [13]. The surface reconstruction also has an effect on the lattice constant, affecting its value. Because nanocrystals have a high surface area to volume ratio, this shift in the lattice constant value is very important [14]. This change in the lattice constant indicates that alternative ions entered the crystal structure in a substitution or interstitial manner between the iron ions, resulting in lattice widening and a decrease in density, which may also be attributed to extra impurities [15].

*3.2. SEM and EDX Analysis:*

Samples of the prepared ferrite ($Cu_xCe_{0.3-x}Ni_{0.7}Fe_2O_4$) were photographed using the Field emission scanning electron microscopy (FE-SEM) technique to clearly characterize the nature of the surface and shape of the particles, as well as their rate of grain size. Figure (2) demonstrates that the material is truly in the nanoscale range. The nanoparticles of the specified material are spherical or semi-spherical in form, with some groupings or agglomerations. There are also spaces between the conglomerates or agglomerations in the area, and these holes indicate the porous quality of the substance's surface, which is required for gas adsorption [16]. The existence of holes is caused by the presence of impurities, which raises the value of the crystal lattice constant, increasing the specific surface area with respect to the compound's volume. The above results in the formation of a porous structure, which improves the sensor's reaction to the test gas [17].

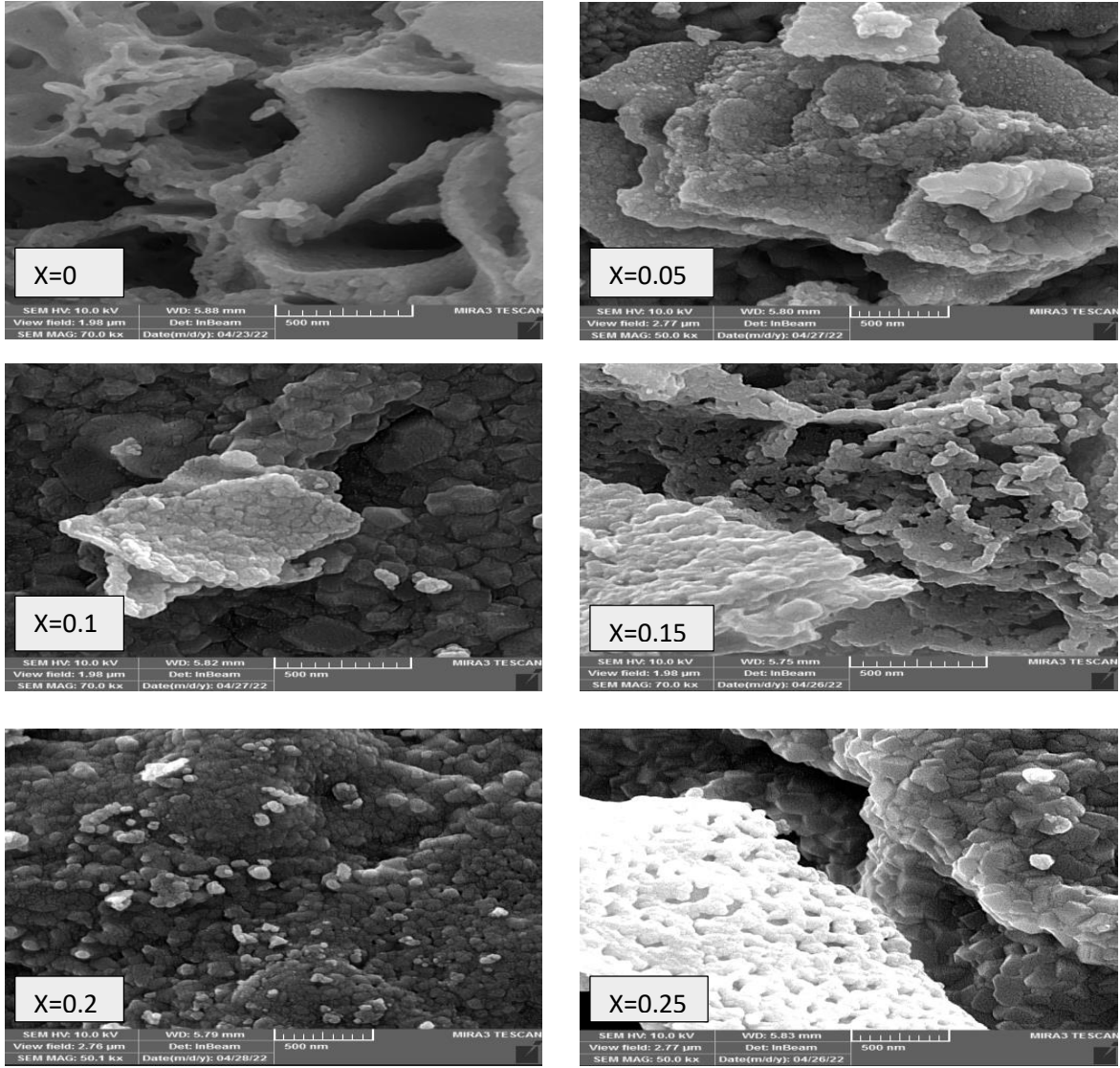

Figure (2). FE-SEM images of the $Cu_xCe_{0.3-x}Ni_{0.7}Fe_2O_4$ ferrite nanocomposite samples.

The energy-dispersive X-ray spectroscopy (EDS) used in conjunction with the emitting field scanning electron microscope (FE-SEM) to confirm the presence of the elements of the prepared compound ($Cu_xCe_{0.3-x}Ni_{0.7}Fe_2O_4$), as shown in Figure (3), shows that all of the elements of the two prepared compounds appear and that all samples are pure and free of impurities.

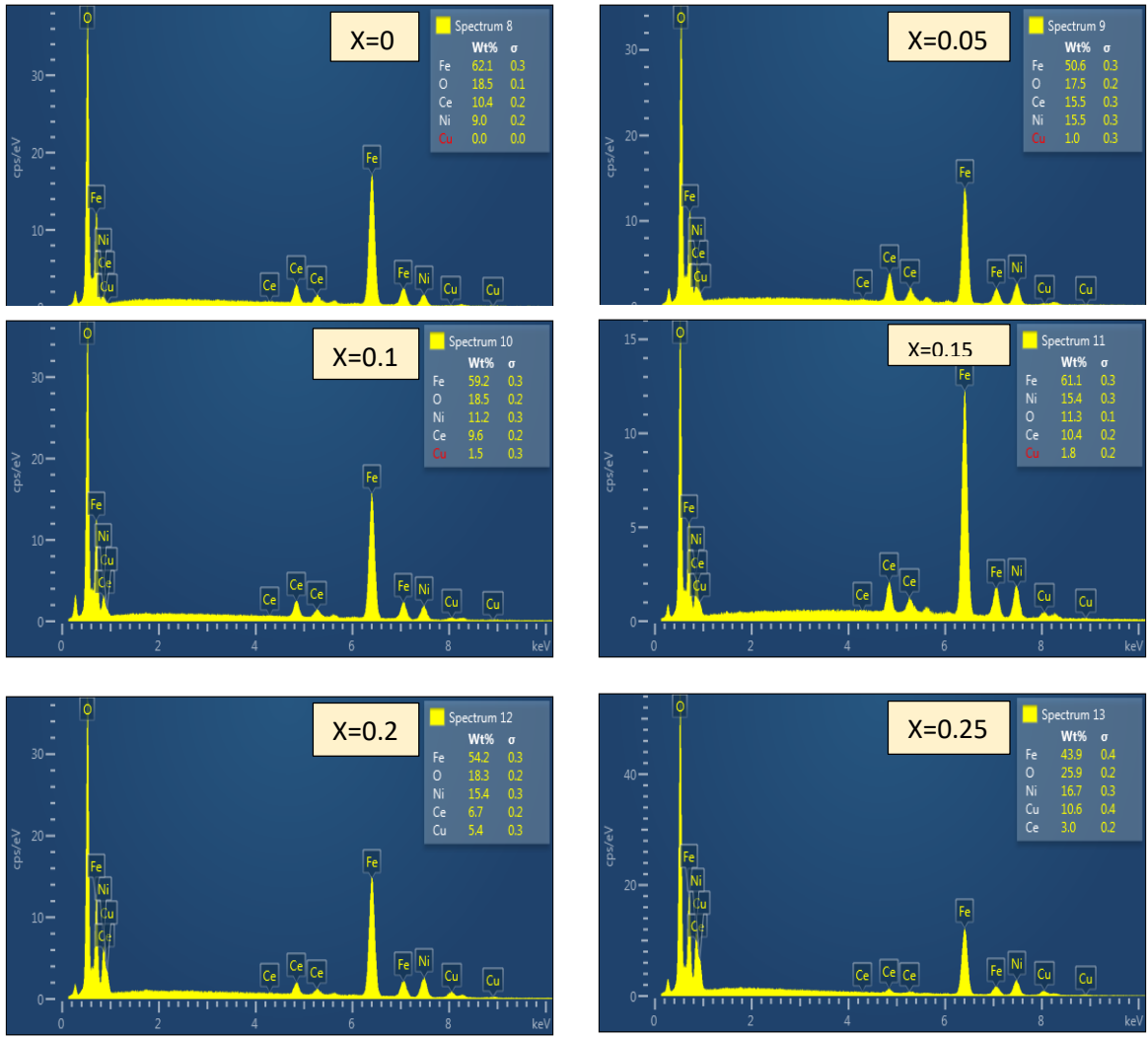

Figure (3). EDS images of the ferrite nanocomposite samples $Cu_xCe_{0.3-x}Ni_{0.7}Fe_2O_4$

### 3.3. Sensing properties:

Equation (2) [18] was used to calculate the sensitivity of $Cu_xCe_{0.3-x}Ni_{0.7}Fe_2O_4$ samples to nitrogen dioxide gas ($NO_2$), and the results indicated that the sensitivity to $NO_2$ gas changes as the operating temperature changes, as illustrated in Figure (4).

$$S = (|R_a - R_g|/R_a)*100\% \quad \ldots\ldots\ldots\ldots\ldots\ldots\ldots\ldots\ldots\ldots\ldots\ldots\ldots\ldots\ldots\ldots\ldots \quad (2)$$

Where: Where: $R_a$ the sensor model's electrical resistance in air, $R_g$ the gas-sensitive model's electrical resistance.

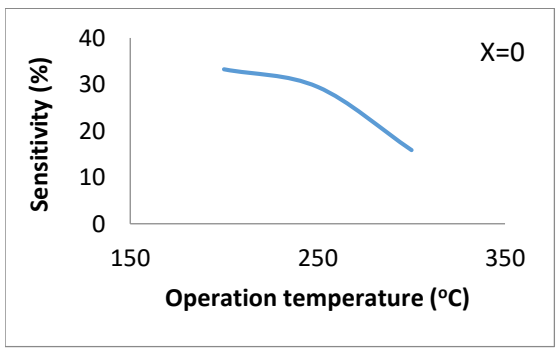
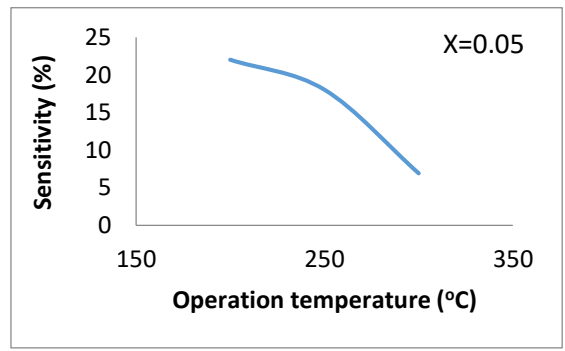

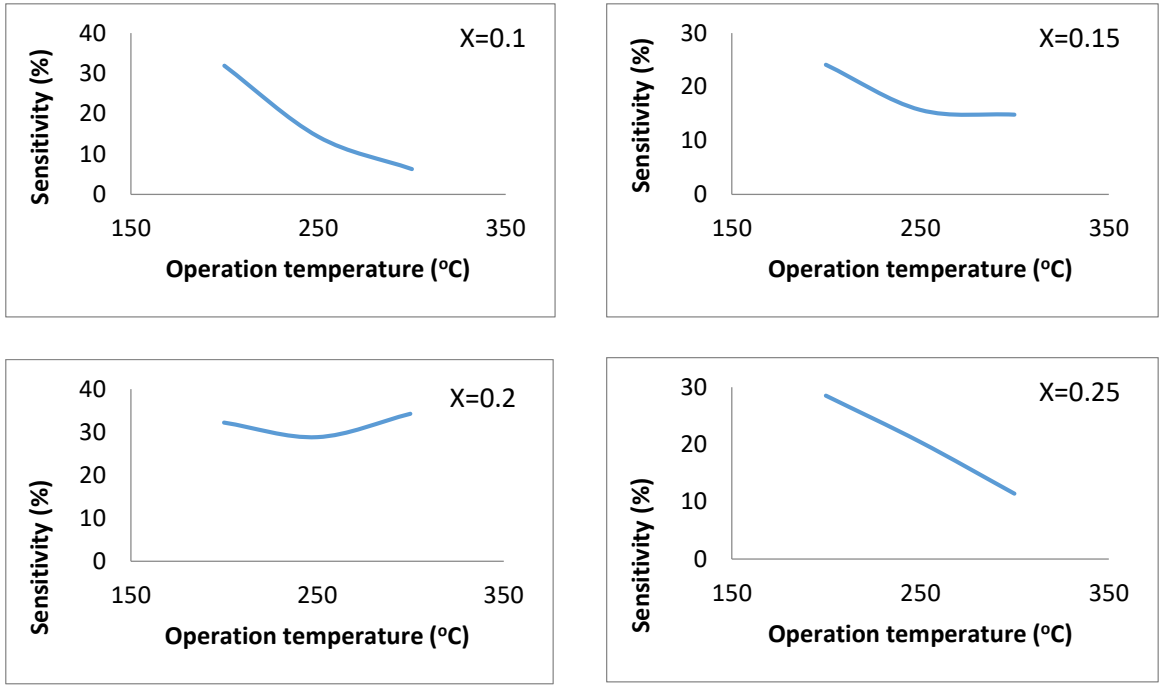

Figure (4). The sensitivity relationship to the operating temperature of the $Cu_xCe_{0.3-x}Ni_{0.7}Fe_2O_4$ samples.

Table (3) shows the highest sensitivity values for the samples of the prepared compound, and it is noted that the highest sensitivity value was at 300 °C when Cu Content is 0.20 mole.

Table (3). The highest sensitivity values of the $NO_2$ gas for $Cu_xCe_{0.3-x}Ni_{0.7}Fe_2O_4$ nanoparticles.

| Cu Content (mole) | Operating temperature (°C) | Highest sensitivity value (%) |
|---|---|---|
| 0 | 200 | 33.2197615 |
| 0.05 | 200 | 22.02247191 |
| 0.1 | 200 | 31.89987163 |
| 0.15 | 200 | 24.0987984 |
| 0.2 | 300 | 34.28285857 |
| 0.25 | 200 | 28.50678733 |

The sensitivity of $Cu_xCe_{0.3-x}Ni_{0.7}Fe_2O_4$ samples to oxidizing nitrogen dioxide ($NO_2$) was investigated. The samples demonstrated an acceptable response to the aforementioned gas, allowing it to be used in various applications. The voltage barrier between the surfaces of molecules is considerable in oxidizing gases, resulting in more resistance to the flow of negative charge carriers, which are electrons, than in reducing gases [19]. The results reveal that the activation process of the ferrite nanoparticles $NiFe_2O_4$ raises the value of the crystal lattice constant, resulting in a porous structure that raises the specific surface area of the compound and, as a result, raises the sensor's sensitivity to the test gas [20].

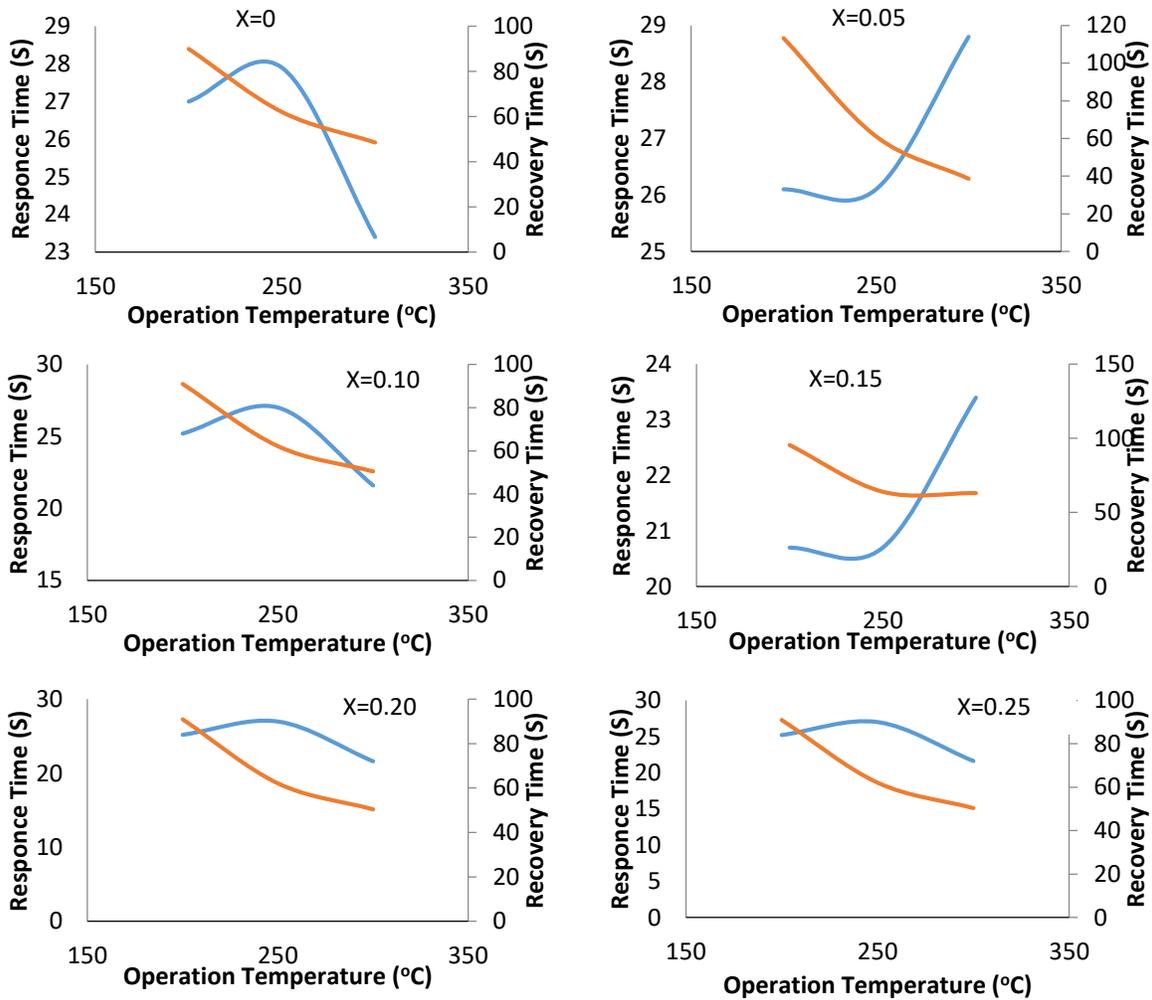

Figure (5). Relationship of response time and recovery time of $NO_2$ gas at operating temperature for ferrite nanoparticles $Cu_xCe_{0.3-x}Ni_{0.7}Fe_2O_4$. Where the red line represents the response time, while the blue line represents the recovery time

Figure (5) illustrates that due to the impurities' influence on the lattice constant value, the addition of impurities to each new ion in specific proportions influences both the reaction time and the recovery time [21]. The reaction time and recovery time are influenced by the speed of the interaction between oxygen atoms in the environment and the gas atoms to be detected with atoms of the granular surface of the sensor material, as well as the geometric form of the columns or electrodes. In turn, these two parameters are affected by the operating temperature as an external, readily controllable component, as well as the structure of the sample [22]. Because there are times when the same sample has a faster reaction and recovery time, the samples that best fit the required application are picked.

Table (4) ) shows that the minimum response time for $NO_2$ gas for samples of the prepared compound was at an operating temperature of 200 °C, while the minimum recovery time was at an operating temperature of 300 °C.

Table (4): Minimum response time and recovery time for ferrite nanoparticles $Cu_xCe_{0.3-x}Ni_{0.7}Fe_2O_4$ samples to $NO_2$ gas.

| Content (mole) | Minimum response time (sec) | Operating temperature (°C) | Minimum recovery time (sec) | Operating temperature (°C) |
|---|---|---|---|---|
| 0 | 24.3 | 200, 250 | 45.9 | 300 |
| 0.05 | 26.1 | 300 | 38.7 | 300 |
| 0.1 | 21.6 | 300 | 50.4 | 300 |
| 0.15 | 20.7 | 200 | 63 | 300 |
| 0.2 | 21.6 | 200 | 46.8 | 300 |
| 0.25 | 23.4 | 300 | 48.6 | 300 |

## 4. Conclusion:

The auto-combustion sol-gel approach was utilized to effectively produce nickel, cerium, and copper ferrite single-phase nanoparticles ($Cu_xCe_{0.3-x}Ni_{0.7}Fe_2O_4$). The XRD and FE-SEM examinations of the manufactured chemical samples indicated that it is a ferrite nanoparticle of a kind of spinel with a multi-crystalline face-centered cubic (FCC) structure and is monophasic. This increased the specific surface area of the compound, and doping the resulting compound with aluminum improved its crystal lattice constant. The samples made for the compound demonstrated an acceptable sensitivity to $NO_2$ gas at various operating temperatures, with the samples recording the highest sensitivity to it at an operating temperature of 300 °C, and the samples' minimum response time was at 200 °C, while the minimum recovery time was at 300 °C.